\documentclass[12pt]{article}

\usepackage{times}
\usepackage{epsfig,amsmath}
\usepackage{subfigure}
\usepackage{graphicx}
\usepackage{color}
\usepackage{epstopdf}
\usepackage{cite}

\newenvironment{sciabstract}{%
\begin{quote} \bf}
{\end{quote}}

\newcounter{lastnote}

\title{128 Identical Quantum Sources Integrated on\\
a Single Silica Chip}

\author
{Ruo-Jing Ren,$^{1,2}$ Jun Gao,$^{1,2}$ Wen-Hao Zhou,$^{1,2}$ Zhi-Qiang Jiao,$^{1,2}$ Lu-Feng Qiao,$^{1,2}$\\
Xiao-Wei Wang,$^{1,2}$ and Xian-Min Jin$^{1,2,\ast}$\\
\\
\normalsize{$^1$Center for Integrated Quantum Information Technologies (IQIT), School of Physics}\\ 
\normalsize{and Astronomy and State Key Laboratory of Advanced Optical Communication Systems}\\
\normalsize{and Networks, Shanghai Jiao Tong University, Shanghai 200240, China}\\
\normalsize{$^2$CAS Center for Excellence and Synergetic Innovation Center in Quantum Information and}\\
\normalsize{Quantum Physics, University of Science and Technology of China, Hefei, Anhui 230026, China}\\
\normalsize{$^\ast$E-mail: xianmin.jin@sjtu.edu.cn}\\
}

\date{}

\begin{document}
\baselineskip24pt

\maketitle

\begin{sciabstract}

Quantum technology is playing an increasingly important role due to the intrinsic parallel processing capabilities endorsed by quantum superposition, exceeding upper limits of classical performances in diverse fields. Integrated photonic chip offers an elegant way to construct large-scale quantum systems in a physically scalable fashion, however, nonuniformity of quantum sources prevents all the elements from being connected coherently for exponentially increasing Hilbert space. Here, we experimentally demonstrate 128 identical quantum sources integrated on a single silica chip. By actively controlling the light-matter interaction in femtosecond laser direct writing, we are able to unify the properties of waveguides comprehensively and therefore the spontaneous four-wave mixing process for quantum sources. We verify the indistinguishability of the on-chip sources by a series of heralded two-source Hong-Ou-Mandel interference, with all the dip visibilities above 90$\%$. In addition, the brightness of the sources is found easily reaching $MHz$ and being applicable to both discrete-variable and continuous-variable platform, showing either clear anti-bunching feature or large squeezing parameter under different pumping regimes. The demonstrated scalability and uniformity of quantum sources, together with integrated photonic network and detection, will enable large-scale all-on-chip quantum processors for real-life applications.
\end{sciabstract}

Over the last few decades, harnessing quantum theories like quantum superposition and Heisenberg's uncertainty principle has led to many quantum enhanced technologies, such as quantum cryptography\cite{GisinRMP}, quantum computating\cite{QC} and quantum metrology\cite{Giovannetti2004}, demonstrating superior capabilities beyond their classical counterparts. Massive efforts have been made in encoding qubit\cite{Eisaman2011} in photon, atom, ion, NV center and quantum dot. Among these physical implementations, photon is a robust and low-noise carrier since it does not easily couple to environment. It is therefore not necessary to keep photonic quantum systems in an extreme environment with ultra-low temperature and high vacuum, which makes quantum technologies run at ambient conditions very appealing for practical applications\cite{OBrien2009}. Exploiting micro-machining processes, integrated quantum photonics provides a feasible platform for generation, manipulation, and detection of optical quantum states by confining light inside waveguide circuits, which further facilitates building large-scale quantum systems in a physically-scalable fashion for practical applications\cite{Wang2020}.  

Scalable generation and high-visibility interference of independent quantum sources is crucial to scale up quantum systems coherently. Efforts to build quantum sources have focused mainly on the following two tracks. The first one relies on nonlinear processes such as spontaneous parametric down-conversion (SPDC)\cite{SPDC} or spontaneous four-wave mixing (SFWM) process\cite{SFWM}, and the other is based on solid-state single emitters such as quantum dots\cite{Qdot}. While solid-state single emitter possesses high brightness and excitation rate, it is very challenging to interface the emitted photons into quantum circuits and very difficulty in finding indistinguishable independent emitters for interconnection\cite{Solidstate}. In contrast, the nonlinear process in the waveguides of a photonic chip can be directly employed to generate quantum sources that support mass manufacturing and are free of additional insertion loss to quantum circuits in a single chip\cite{Lightrev}. The remained long-standing challenge is that the quantum sources from each mass-manufacture waveguide are usually independent and nonidentical, and therefore can not be coherently connected to form a whole system with exponentially increasing Hilbert quantum state space.

The inner mechanism is highly related to the nonlinear process. In $\chi^{(3)}$ nonlinear interaction, the material absorbs two photons from the pump wave, and generates signal and idler photon pairs, following energy and momentum conservation, where birefringence induced phase matching condition is fulfilled by
\begin{equation}
\Delta k=\frac{2\left[n\left(\omega_{p}\right)+\Delta n\right] \omega_{p}}{c}-\frac{n\left(\omega_{s}\right) \omega_{s}}{c}-\frac{n\left(\omega_{i}\right) \omega_{i}}{c}=0
\end{equation}
Here the birefringence $\Delta n$ dominates the phase matching condition. The progresses have been made in developing integrated quantum sources by using femtosecond laser writing\cite{Spring2013}, UV-laser writing\cite{Spring2017} and silicon photonics\cite{Silicon}. Such prototypes have verified the availability and performances like high brightness, high purity and low propagation loss with a maximum integration number up to 18 in a single chip. The challenge that hampers the source towards large-scale integration is the sensitivity of $\Delta n$ during fabrication\cite{Spring2017}. Even a small variance would render wavelength shifts of the generated photon pairs. 

As is shown in Fig.1(a), the shifts of virtual energy level induced by small variances of $\Delta n$ lead to the shifted wavelengths of the generated photon pairs. Suppose building active multiplexed single-photon sources or generating high-dimensional entanglements from such integrated source array, the overall generated state $\rho=\sum_{n=1}^{\infty} \rho_{fn}$ will become a mixed state containing ensembles of different spectra. If we can eliminate the fluctuation of the virtual energy level by birefringence engineering, as Fig.1(b) shows, we will be able to lock the SFWM process to produce hundreds of identical photon pairs, thus enabling large-scale integration of such sources for quantum network and high-dimensional quantum computation.

We fabricate waveguides in a fused silica substrate by femtosecond laser direct writing\cite{Fab} as depicted in Fig.1(c). The substrate is perpendicular to the femtosecond laser whose focus can introduce permanent refractive index change. The substrate is translated by an air-bearing high-precision 3D positioning stages. To form a waveguide which can generate non-classical photon pairs and meanwhile avoid spontaneous Raman scattering noise, a sufficiently large and stable birefringence engineering ability is needed. Over the last decades, various laser writing methods have been adopted for birefringence engineering, such as introducing damage tracks\cite{Damage} and multiple pass writing\cite{Multipass}. However, these approaches lead to great photon loss which hinder future all-on-chip mass-integration. By carefully scanning the laser writing parameters like pulse energy, writing velocity, repetition rate and focusing condition, we manage to fabricate a high-quality SFWM source waveguide in single track without extra modifications (see Methods for fabrication details).

To reach the ultimate goal of mass-manufacture and fast prototyping quantum source array, stability is another crucial requirement. Our micro-machining system integrates many sensors to monitor the status of the fabrication laser, like temperature, humility, laser power and repetition rate. We further install steering mirror with feedback system to lock the beam pointing inside the sample. After all these efforts, the micro-machining system can work in a stable and accurate way. The 128 uniform SFWM source waveguides can be inscribed on a 20$mm$$\times$20$mm$$\times$1$mm$ fused silica substrate in less than one hour due to mask-free advantage of femtosecond laser direct writing, and in principle, can be extended to serval thousands using either larger size of substrate or 3D architecture.

We inject mode-locked 780$nm$ femtosecond pump pulses along the slow axis (vertical polarization) of the waveguide, then the polarization of the generated correlated photon pairs, namely signal and idler photons are along the fast axis (horizontal polarization). The schematic of the experimental setup is shown in Fig1.(d), the residual pump light and generated photon pair are sophisticatedly filtered (see details in Methods), and the signal and idler photons are separated into two different spatial modes by a dichroic mirror, where the idler photons transmit while the signal photons are reflected. The photon pairs are coupled into single mode fiber with a collection efficiency over 80$\%$, and then detected by avalanche photodiodes and all the coincidence counts are recorded by a homemade multi-channel coincidence module. The cross correlation $g_{si}^2(0)$ is measured to be 160.49$\pm$3.05 under 10$mW$ pumping, which is a strong evidence of non-classical correlation. We can conveniently switch the measurement setup between different channels and further perform Hong-Ou-Mandel interference with chosen pairs.

To confirm the uniformity of all 128 SFWM sources, we experimentally measure all the spectra of the signal and idler photons by using a single-photon-sensitive spectrometer with a resolution of 0.2$nm$. As shown in Fig.2(a), the average central wavelength of the signal photons is 732.5$\pm$0.4$nm$ while that of the idler photons is 833.4$\pm$1.1$nm$. We present the wavelength fluctuation and derived perturbation of birefringence based on our experiment parameters (see Methods) in Fig.2(b) and (c). When the perturbation of the birefringence $\Delta n$ reaches 5$\%$, the differences of the source array are about 1$nm$, which shows that our micro-machining has locked the birefringence discrepancy well below 5$\%$. All measured 128 pairs of spectra are illustrated in Fig.2(d) and (e). From these results, the measured 128 relatively narrow spectra of the signal photons are identical, revealing a good uniformity of the spectrum property. And for the wider idler photons spectra, despite showing a little deviation from the average central wavelength, are still within acceptable tolerance.

The generated photon pairs can be described by a two-mode squeezed vacuum state\cite{SV},  
\begin{equation}
|\psi\rangle_{SFWM}=\sum_{n=0}^{\infty} c_{n}|n\rangle_{s}|n\rangle_{i}
\end{equation}
Where $n$ is an integer representing different orders, $s$ and $i$ represent signal and idler photon respectively. $c_n=tanh^{n}(r)/{cosh(r)}$ are complex coefficients and $r$ is the squeezing parameter that is related to the pumping power. When the pump power is low, only the first and second order in this formula dominate and the state can serve as a high-quality photon pair or heralded single photon. We scan different pumping powers and verify the corresponding anti-bunching feature by sending the signal photon through the Hanbury-Brown-Twiss interferometer and measure the second­ order anti-correlation function $g_H^{2}(0)=N_{12i}N_{i}/({N_{1i}N_{2i}})$, where the idler photon serves as the trigger. The measured results are presented in Fig.3(a) and the values are all below 0.12, showing a clear anti-bunching feature and suppression of multi-photon emission.

Besides the degrees of freedom such as polarization, path, or orbital angular momentum of single photons, quantum information can also be carried by continuous-variable field of light, which is an important branch of quantum information science\cite{CVRMP}. The quantum light sources required to implement continuous-variable protocols are mostly squeezed light. In the high pumping regime, multi-photon terms in $|\psi\rangle_{SFWM}$ can not be ignored any more and contribute to form a genuine two-mode squeezed vacuum state. In a bulk crystal, a tight focusing condition may lead to strong nonlinear interaction, but the Rayleigh length is usually very small, thus limiting the interaction zone. While the transverse mode confines the light strictly inside the material all along the waveguide, leading to a great enhancement to $\chi^{(3)}$ nonlinear interaction. Our on-chip source array meets these two requirements simultaneously, and therefore can generate cavity-free high squeezing states. In Fig.3(b), we scan a large power range from 10$mW$ to 150$mW$ of the pump laser and record the corresponding coincidence count rates. We can see from the purple line that the coincidence count can easily reach $MHz$ level. As the pump power goes up, the nonlinear interaction in the waveguide increases dramatically, which leads to a large squeezing parameter. The results are also illustrated as the blue line in Fig.3(b), the achieved squeezing parameter can go up to 0.545$\pm$0.01, suggesting our quantum source array as a new platform for both discrete-variable and continuous-variable regions.

Beyond the spectrum uniformity, indistinguishability of quantum particle is a more crucial desired feature for scalable quantum information processing, such as boson sampling\cite{AA2013} and KLM protocol\cite{KLM}, which require the heralded single photons must be pure and identical in terms of all the degree of freedom. We divide 128 SFWM sources into ten groups, and verify the indistinguishability via a series of Hong-Ou-Mandel interference experiments\cite{HOM} using heralded single photons randomly picked from two groups. The idler photons are injected into a $50:50$ fiber beam splitter while the signal photons are directly sent to the single-photon detectors, serving as the triggers. Temporal overlap of the photons are achieved by a linear translation stage with a step size of 0.02$mm$. We measure ten groups of Hong-Ou-Mandel interference and enumerate all the experimental results in Fig.4. From the Hong-Ou-Mandel dips, all the calculated interference visibility as $V=(P(\tau)-P(0))/P(\tau)$ are well beyond 0.9, indicating a great indistinguishability among the arrayed quantum sources.

In summary, we report 128 identical quantum sources integrated on a single silica chip in a scalable fashion. The scale of quantum source array is expanded by an order of magnitude meanwhile showing high-quality single-photon anti-bunching and Hong-Ou-Mandel interference. The uniformity of the source is guaranteed by birefringence engineering associated with nonlinear interaction locking in the fabrication process of femtosecond laser direct writing. Moreover, the arrayed quantum sources are tunable ranging from high-quality photon pairs to strongly squeezed states in different pumping regime, being able to stimulate quantum applications in a large scale and high dimension for both discrete-variable and continuous-variable approaches. 

Future investigation may promptly extend to large-scale driven quantum walk\cite{DQW1,DQW2}, scattershot boson sampling\cite{SBS}, and Gaussian boson sampling\cite{GBS}. Active multiplexing can also be applied to our system to enhance the emission probability or synchronized multi-photon generation rate per clock cycle\cite{Multiplexing}. Combined with on-chip photonic network and detectors\cite{Onchipdet1,Onchipdet2}, the achieved arrayed quantum sources makes it possible to realize the generation, manipulation and measurement on a single silica chip, representing a solid step towards all-on-chip quantum information processing.\\

\subsection*{Acknowledgments.}
The authors thank Jian-Wei Pan for helpful discussions. This work was supported by National Key R\&D Program of China (2019YFA0308700 and 2017YFA0303700); National Natural Science Foundation of China (NSFC) (61734005, 11761141014, 11690033); Science and Technology Commission of Shanghai Municipality (STCSM) (17JC1400403); Shanghai Municipal Education Commission (SMEC) (2017-01-07-00-02-E00049); X.-M. J. acknowledges additional support from a Shanghai talent program.
\\

\subsection*{Methods}
\paragraph*{Fabrication of arrayed identical source:} The laser pulse for fabricating the source array is generated by a regenerative amplifier based on Yb:KGW laser medium, with a pulse duration of 290$fs$, a repetition rate of 1$MHz$ and a central wavelength of 513$nm$. We reshape the femtosecond laser beam into a narrow one with a cylindrical lens, whose focal length is 700$mm$. Then we focus the laser with a 100X (0.70 NA) microscope objective. The waveguides are inscribed by a fixed pulse energy of 260$nJ$ and a constant velocity of 1.268$mm/s$, 75$\mu m$ below the surface of the fused silica substrates (Corning 7980, 20$mm$$\times$20$mm$$\times$1$mm$ with all facets optically polished). The coupling efficiency to standard single-mode fibers is over 80$\%$ without the need for lossy spatial filtering. 

The inscribed 128 waveguides are identical to generate uniform photon pairs and integrated on a single photonic chip. Typically, it is always a difficulty to keep the waveguides identical for large scale, especially up to hundreds of trials. There are some dominating parameters to affect the fabrication process, laser power, writing velocity, polarization, position precision of the translation stages and beam pointing. During the fabrication process, we lock the power of the writing laser less than 0.5$\%$rms over 100 hours under our lab environment. Besides, a feedback system is applied in our system to fix the pointing angle less than 0.5$\mu rad$. A high-precision 3D air-bearing stage keeps the position in a deviation of $\pm$50$nm$, which is a negligible error compared to the depth of 75$\mu m$. All the parameters are maintained uniform to guarantee the stability during the fabrication process. 

\paragraph*{Filtering details:} A vertically polarized pump femtosecond pulse centered at 780$nm$ is injected into the waveguide and the signal and idler photons are both horizontally polarized according to the phase matching condition. The residual pump light and generated photon pair pass a polarization filter (Glan-Taylor prism) and a series of notch filters. The generated photon pairs are then divided by a dichroic mirror. An additional short pass filter as well as a long pass filter are placed in each path accordingly to purify the spectra. To filter out undesired background noises, we further take full advantage of two 12$nm$ tunable bandpass filters. By slightly tuning the angle of the tunable filters, we can select certain spectrum range with maximum correlated photon counts. Additional 1$nm$ bandpass filters are utilized to further purify the frequency-uncorrelated heralded single photons by filtering out the side lobes in the Hong-Ou-Mandel interference experiments.

\paragraph*{Birefringence perturbation:} In our experiment, the pump light wavelength is chosen as 780$nm$ and the measured central wavelengths of the generated photon pairs are 732.5$nm$ and 833.5$nm$ respectively. Following energy and momentum conservation,
\begin{equation}
\Delta \omega=2\omega_{p}-\omega_{s}-\omega_{i}=0
\end{equation}
\begin{equation}
\Delta k=\frac{2\left[n\left(\omega_{p}\right)+\Delta n\right] \omega_{p}}{c}-\frac{n\left(\omega_{s}\right) \omega_{s}}{c}-\frac{n\left(\omega_{i}\right) \omega_{i}}{c}=0
\end{equation}
we can infer that the birefringence of the waveguide is about $6\times10^{-5}$. By introducing perturbation $\eta$ as $\Delta n^{'}=\Delta n\times(1\pm\eta)$, we can speculate the upper and lower bound of the wavelength fluctuations. We scan the perturbation range from 0 to 20$\%$, and draw the results in Fig2.(b) and (c). The results show that the largest fluctuation of the wavelength can reach about $\pm$5$nm$. Even when the perturbation of the birefringence $\Delta n$ is 5$\%$, the wavelength difference of the source array are over 1$nm$. Compared with Fig2.(a), the birefringence discrepancy of the 128 quantum sources are well below 5$\%$, which indicates that our micro-machining has locked the birefringence discrepancy well below 5$\%$, demonstrating good uniformity of the spectrum property. 

\clearpage

\clearpage

\begin{figure}[htbp]
	\centering
	\includegraphics[width=1.0\linewidth]{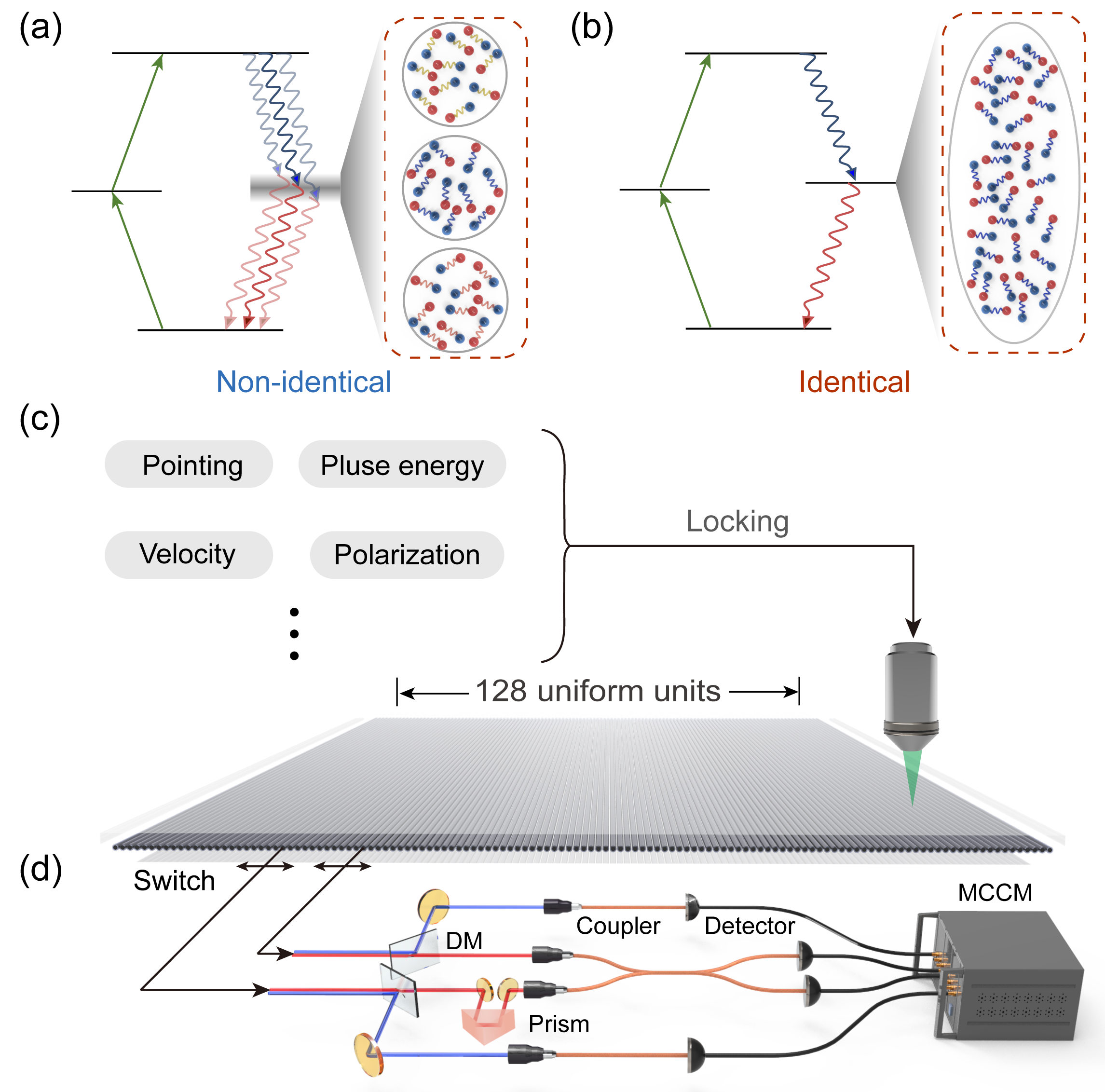}
	\caption{\textbf{Schematic of nonlinear interaction locking and 128 identical SFWM integrated quantum sources.} \textbf{(a)} Energy level diagram of SFWM process. Non-identical photon emission ensembles are induced by virtual energy levels variances. \textbf{(b)} Identical photon emission without virtual energy level variance. \textbf{(c)} Schematic of femtosecond laser direct writing and birefringence engineering with comprehensive fabrication parameters locking. \textbf{(d)} Experimental setup for the characterization of the arrayed quantum sources. The testing apparatus can be easily switched between different groups. DM: dichroic mirror, MCCM: multi-channel coincidence module.}
	\label{f1}
\end{figure}

\clearpage

\begin{figure}[htbp]
	\centering
	\includegraphics[width=0.9\linewidth]{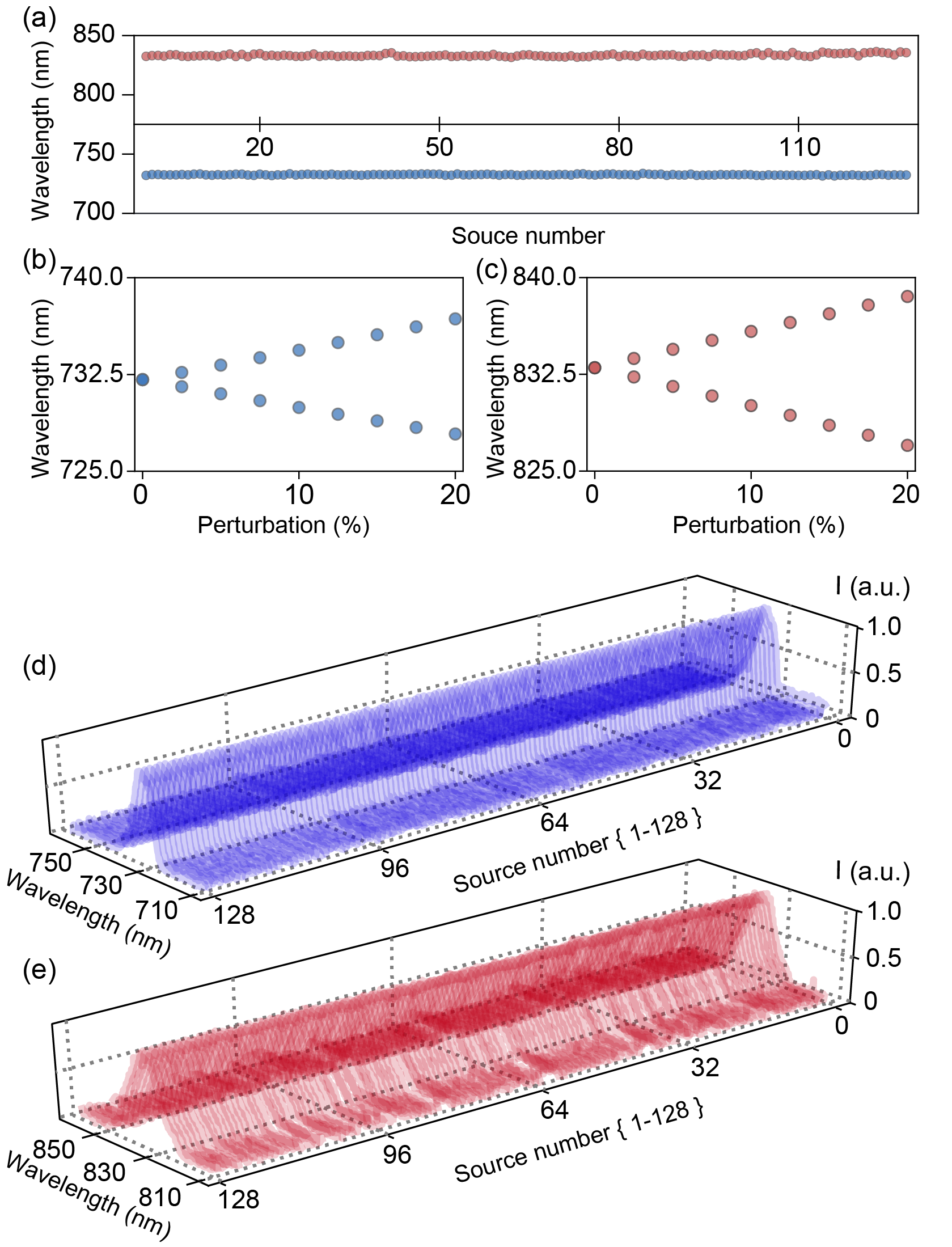}
	\caption{\textbf{Homogeneous Spectra of 128 identical SFWM integrated sources.} \textbf{(a)} Experimentally measured central wavelengths of 128 SFWM integrated sources. Signal photon is depicted by blue dots while idler photon of red dots. \textbf{(b)} Signal wavelength drifting with perturbation of birefringence $\Delta n$. \textbf{(c)} Idler wavelength drifting with perturbation of birefringence $\Delta n$. \textbf{(d)} Measured 128 signal full spectra in blue color. \textbf{(e)} Measured 128 idler full spectra in red color. The intensity of spectrum measurement is normalized to its maximum.}
	\label{f2}
\end{figure}

\clearpage

\begin{figure}[htbp]
	\centering
	\includegraphics[width=1\linewidth]{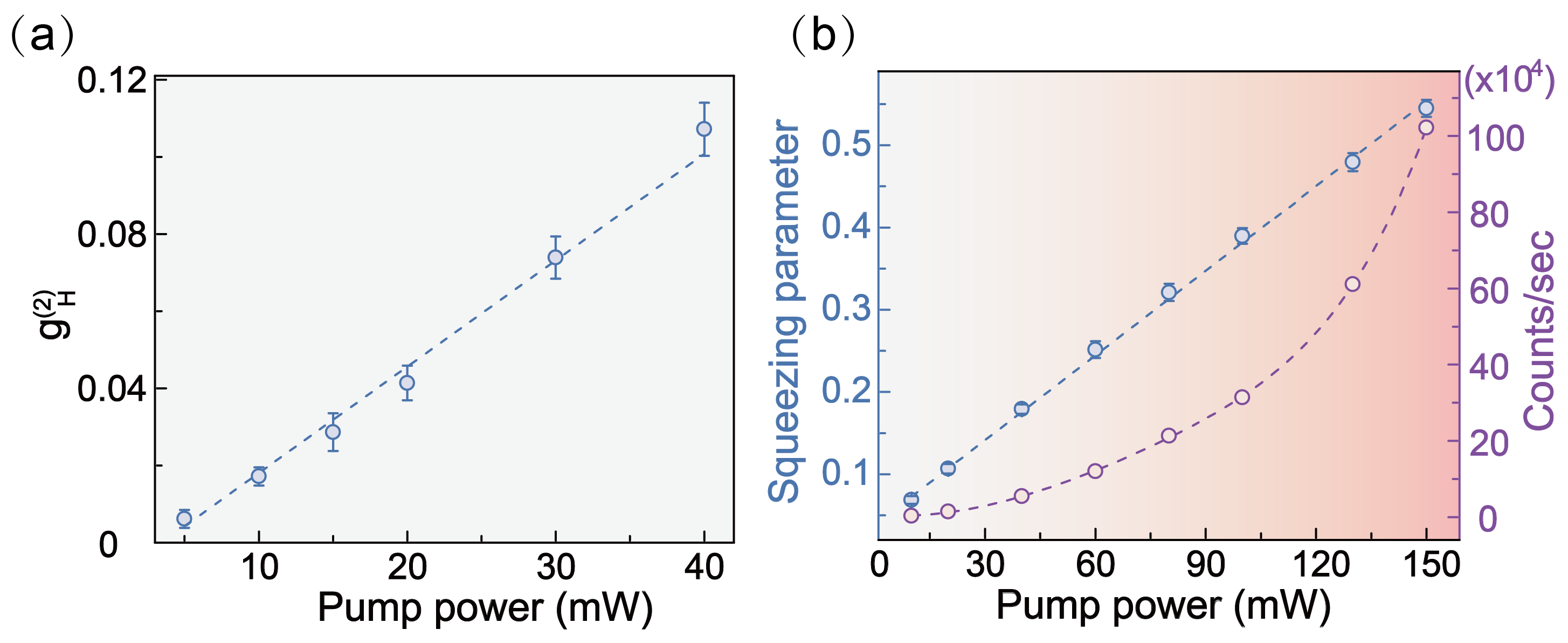}
	\caption{\textbf{Experimental characterization of quantum sources.} \textbf{(a)} Heralded $g_{H}^{2}(0)$ in the low pump regime, the results are given by the Hanbury-Brown-Twiss experiments, indicating a clear anti-bunching feature. \textbf{(b)} While scanning pump power range from 10$mW$ to 150$mW$, both the squeezing parameter and coincidence counts are depicted with blue and purple colors. Some error bars are too tiny to be visible.}
	\label{f3}
\end{figure}

\clearpage

\begin{figure}[htbp]
	\centering
	\includegraphics[width=1.0\linewidth]{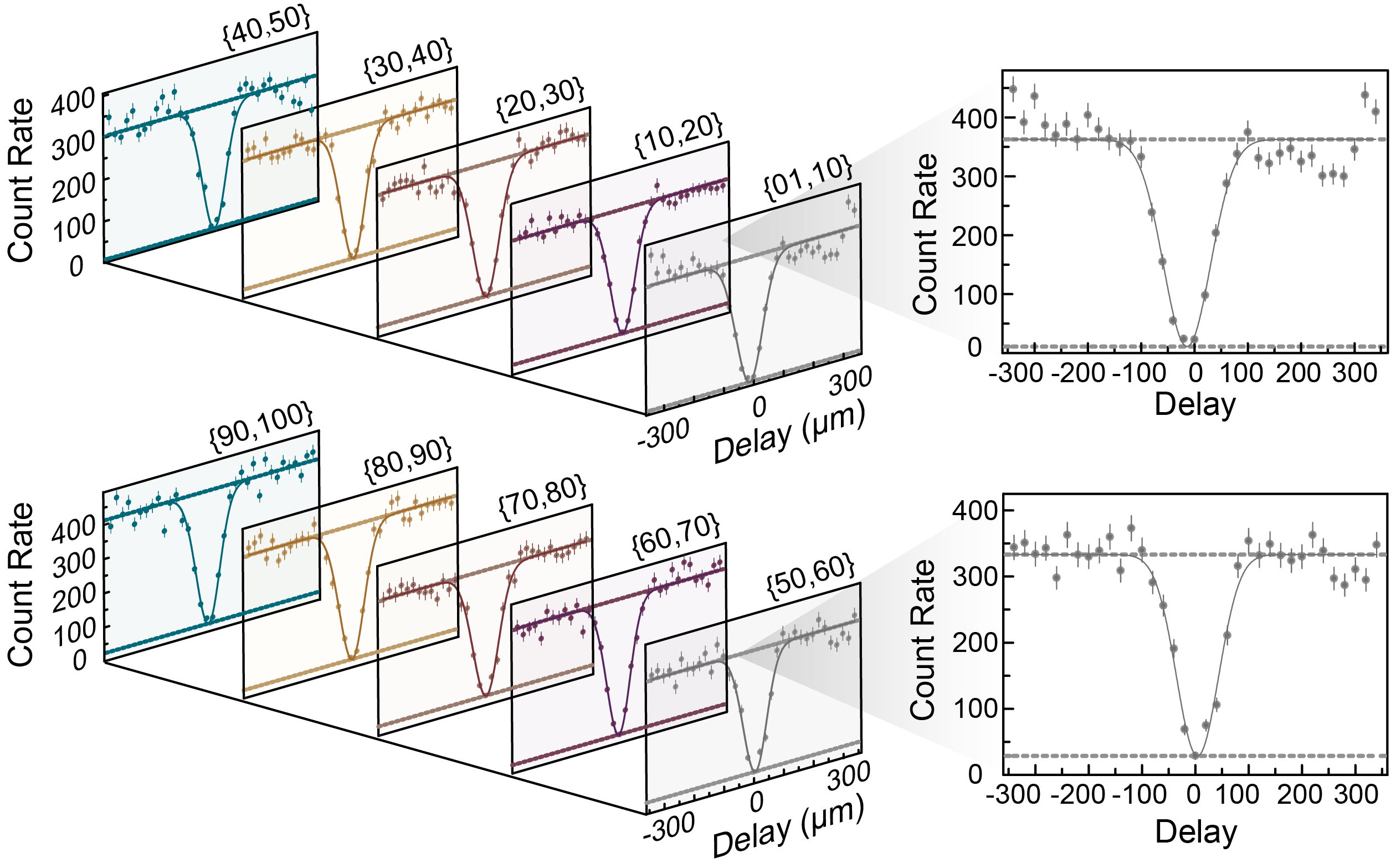}
	\caption{\textbf{Interconnected test of Hong-Ou-Mandel interference.} With the experimental apparatus plotted in Fig.1\textbf{(d)}, a series of Hong-Ou-Mandel interferences are conducted with visibilities all beyond 0.9, showing a good indistinguishability and uniformity of the arrayed quantum sources. The experimental data of \{01,10\} and \{50,60\} are enlarged to show more details. } 
		\label{f4}
\end{figure}

\clearpage

\end{document}